\documentclass[a4paper,12pt]{article}
\usepackage[margin=1.0in]{geometry}
\usepackage[utf8]{inputenc}
\usepackage [english] {babel}
\usepackage{amsmath,amssymb}
\numberwithin{equation}{section}
\usepackage{setspace}
\usepackage{textcomp}
\usepackage[numbers]{natbib}
\usepackage{hyperref}

\def \sec{\begin{section}}
\def \esec{\end{section}}
\def \beq{\begin{equation}}
\def \eeq{\end{equation}}

\def \la {\lambda}
\def \La {\Lambda}
\def \om {\omega}
\def \Om {\Omega}

\def \ep {\epsilon}

\def \pr {\partial}
\def \ra {\rightarrow}

\def \th {\theta}
\DeclareMathOperator*{\res}{res}
\DeclareMathOperator*{\tr}{tr}

\renewcommand\Re{\operatorname{Re}}
\renewcommand\Im{\operatorname{Im}}
\newcommand\const{\operatorname{const}}

\def \l {\left(}
\def \r {\right)}

\begin{document}

\begin{flushright}
 \small{ITEP-TH-19/14}\\
\end{flushright}

\begin{center}
{  \Large \bf RG -Whitham dynamics and complex Hamiltonian systems }
\end{center}
\vspace{1mm}

\begin{center}
{\large
   A.~Gorsky$^{\,2,3}$\footnote{gorsky@itep.ru},  A.~Milekhin$^{\,1,2,3}$\footnote{milekhin@itep.ru}\\ }
\vspace{3mm}
$^1$Institute for Theoretical and Experimental Physics, B.Cheryomushkinskaya 25, Moscow 117218, Russia \\
$^2$Moscow Institute of Physics and Technology, Dolgoprudny 141700, Russia \\
$^3$ Institute for Information Transmission Problems, B. Karetnyi 15, Moscow 127051, Russia
\end{center}

\vspace{0.5cm}

\begin{abstract}
Inspired by the Seiberg-Witten exact solution, we consider some aspects of the Hamiltonian dynamics with the 
complexified phase space focusing at the renormalization group(RG)-like Whitham behavior. We show that at the Argyres-Douglas(AD) point the number of 
degrees of freedom in Hamiltonian system effectively reduces and argue that anomalous dimensions  at AD point coincide with the  
Berry indexes in classical mechanics.
In the framework of  Whitham dynamics AD point turns out to be a fixed point.
We demonstrate that  recently discovered Dunne-\"Unsal relation in quantum mechanics
relevant for the exact quantization condition
exactly coincides with the Whitham equation of motion in the $\Omega$ - deformed theory.  
\end{abstract}

\tableofcontents

\section{Introduction}

The holomorphic and complex Hamiltonian systems  attract now the substantial interest partially motivated by their 
appearance in the Seiberg-Witten solution to the $N=2$ SUSY YM theories \cite{sw}. 
They have some essential differences in comparison with the real case mainly due to the nontrivial 
topology of the fixed energy Riemann surfaces in the phase space. Another subtle issue concerns the choice of the quantization condition which is not unique.

The very idea of our consideration is simple - to use some physical intuition developed
in the framework of the SUSY gauge theories and apply it back to complex or holomorphic Hamiltonian systems
which are under the carpet. The nontrivial phenomena at the gauge side have interesting
manifestations in the dynamical systems with finite number degrees of freedom. There are a few different dynamical systems in SUSY gauge theory framework. In the $N=2$ case one can define a pair of the dynamical  systems related with each other in a well defined manner (see \cite{sw_int_rev} for review). The second Whitham-like Hamiltonian system \cite{krichever} is defined on the moduli space of the first Hamiltonian system. Note that there is no need for the first system to be integrable
while the Whitham system is certainly integrable. It can be considered as the RG flow in the field theory framework \cite{Whitham98}. 

One
more dynamical system can be defined upon the deformation to $N=1$ SUSY where the chiral ring relation plays
the role of its energy level. In this case one deals with the Dijkgraaf-Vafa matrix model \cite{dv0206,dv0207} in the large N limit. It is known that matrix models in the large N limit give rise to one-dimensional mechanical system, with the loop equation playing the role
of energy conservation and 1-point resolvent playing the role of action differential $pdq$. The degrees of freedom  in all cases
can be attributed to the brane coordinates in the different dimensions and mutual coexistence  of the
dynamical systems plays the role of the consistency condition of the whole brane configuration.
We shall not use heavily the SUSY results but restrict ourselves only by application of a few important issues
inherited from the gauge theory side to the Hamiltonian systems with the finite number
degrees of freedom. Namely we shall investigate the role of the RG-flows, anomalous dimensions at AD points
and condensates in the context of the classical and quantum mechanics.

First we shall focus at the behavior of the dynamical system near the AD point. It is interesting
due to the following reason. It was shown in \cite{condensate} that the AD point in the softly broken $N=2$ theory corresponds
to the point in the parameter space where the deconfinement phase transition occurs. The field theory analysis is performed into two steps. First the AD point at the moduli space of $N=2$ SUSY YM theory gets identified and than the vanishing of the monopole condensate which is the order parameter is proved upon the perturbation. The consideration in the complex classical mechanics is parallel to the field theory  therefore the first step involves the explanation of the AD point before any perturbation. We argue that the number of degrees of freedom at AD point gets effectively reduced which is the key feature of the AD point in classical mechanics.
Moreover we can identify the analog of the critical indices at AD point in  Hamiltonian system as   the Berry indexes relevant for the  critical
 behavior near caustics. Also, we propose a definition for a "correlation length" for a mechanical system so that corresponding
 anomalous dimensions coincide with the field-theoretical ones. From the Whitham evolution viewpoint the AD point is the fixed point. However the second step concerning the perturbation and identification of the condensates is more complicated and we shall restrict
ourselves by the few conjectures. Note that the previous discussion of the Hamiltonian interpretation of the AD points can be found in \cite{smilga} however that paper was focused at another aspects of the problem.

Quantization of complex quantum mechanical systems is more subtle and we consider the
role of the Whitham dynamics in this problem. The progress in this direction 
concerns the attempt to formulate the exact energy quantization condition which involves the
non-perturbative instanton corrections. It turns out that at least in the simplest examples 
\cite{zinn04} the exact quantization condition involves only two functions. Later the
relation between these two functions has been found \cite{dunne13}.
We shall argue  that the Dunne--\"Unsal relation \cite{dunne13} which supplements
the Jentschura--Zinn-Justin quantization
condition \cite{zinn04} can be identified as the equation of motion in the Whitham theory.
To this end we derive the Whitham equations in the presence of $\Om$-deformation, which has not been done in a 
literature before.

The paper is organized as follows. Whitham dynamics is briefly reviewed in Section 2. In Section 3 we shall consider the different aspects of the
AD points in the classical mechanics. Section 4 is devoted to the clarification of the 
role of the Dunne--\"Unsal relation and to the derivation of Whitham equations in the $\Om$-deformed theory. Also, we discuss 
various quantization conditions for complex systems and elucidate the role of the curve of marginal stability. 
The key findings of the paper are summarized in the Conclusion.
In the Appendix we show how the Bethe ansatz equations are modified by the higher Whitham times.

\section{Whitham hierarchy}
\subsection{Generalities}
\label{sec:geom}
Let us define some notations which will be used later, for a nice review see \cite{wh_review}.
Hyper-elliptic curve is defined by
\begin{equation}
y^2=P_{2N}(x)
\end{equation}
where $P_{2N}(x)$ - is polynomial of degree $2N$ -  below, we will be mostly concerned with this particular case. 
There are $2g=2N-2$ cycles $A_i, B_i, i=1...g$ which can be chosen  as follows $(A_i,A_j)=0, (B_i,B_j)=0, (A_i,B_j)=\delta_{ij}$.
For genus g hyper-elliptic curve there are exactly g holomorphic \textit{abelian-differentials of the first kind} $\om_k$:
\begin{equation}
\oint_{A_j} \om_k = \delta_{j k}
\end{equation}
which are linear combinations of $dx/y,...,x^{g-1} dx/y$.
Period matrix is given by:
\begin{equation}
\oint_{B_j} \om_k = \tau_{j k}
\end{equation}

Define  $d \Om_j$ - meromorphic \textit{abelian differential of the second-kind} by the following requirements:

normalization:
\begin{equation}
\oint_{A_k} d \Om_j = 0
\end{equation}
and behavior  near some point(puncture):
\begin{equation}
d \Om_j \approx (\xi^{-j-1}+O(1))d\xi, \ \xi \ra 0
\end{equation}
$d \Om_0$ is actually \textit{abelian differential of the third kind} - with two simple poles with residues $+1,-1$.

Below we will use Riemann bilinear identity  for the pair of  meromorphic differentials $\tilde \om_1,\tilde \om_2$:
\begin{equation}
\sum_{j=1}^g \l \oint_{A_j} \tilde  \om_1 \oint_{B_j} \tilde \om_2 -  \oint_{B_j} \tilde  \om_1 \oint_{A_j} \tilde \om_2 \r
= 2 \pi i \sum_{poles} (d^{-1} \tilde \om_1) \tilde \om_2
\end{equation}
however sometimes it is more convenient to work with non-normalized differentials $dv_k=x^k dx/y$:
\begin{equation}
\om_{kl}=\oint_{A_l} dv_k
\end{equation}
\begin{equation}
\om_{lk}^D=\oint_{B_l} dv_k
\end{equation}

Recall now  some general facts concerning Whitham dynamics.
In classical mechanics, action variables $a_i$ are independent of time. However, sometimes it is interesting to
consider a bit different situation when some parameters of the system  become
\textit{adiabatically dependent} on times. Then the well-known adiabatic theorem states that unlike other possible
integrals of motion, $a_i$ are still independent (with exponential accuracy) on times.

While considering finite-gap solutions to the  integrable system one deals with a spectral curve and a tau-function
\begin{equation}
\label{eq:tau}
\tau=\theta(\sum_j t_j \vec{U}^{(j)}),
\end{equation}
$\theta(z|\tau)$ - is a conventional theta-function,
\begin{equation}
\theta(\vec z|\tau) = \sum_{\vec k} \exp( (\vec z, \vec k) + \pi i (\vec k, \tau \vec k))
\end{equation}
If one introduces "slow"(Whitham) times $t_i=\epsilon T_i, \epsilon \ra 0$,
Whitham hierarchy equations tell us how moduli can be slowly varied provided (\ref{eq:tau}) still gives the solution
to the leading order in $\epsilon$ \cite{babelon},\cite{krichever}.
These equations have zero-curvature form \cite{krichever}
\begin{equation}
\label{eq:whbig}
\cfrac{\pr d \Om_i}{\pr T_j}=\cfrac{\pr d \Om_j}{\pr T_i}
\end{equation}
This guarantees the existence of $dS$ such that
\begin{equation}
\label{eq:whprep}
\cfrac{\pr dS}{\pr T_i} = d \Om_i
\end{equation}
which results in the adiabatic theorem:
\begin{equation}
\label{eq:adiab}
 \cfrac{\pr a_i}{\pr T_j}=0
\end{equation}
The full Whitham-Krichever hierarchy (\ref{eq:whbig}) has a variety of solutions. Every $dS$ satisfying (\ref{eq:whprep})
generates some solution.

Here we have to stop and make one comment concerning the closed Toda chain-case. The Hamiltonian is given by
\begin{equation}
 H=\sum_{i=1}^{N+1} \cfrac{p_i^2}{2} + \La \sum_{i} \exp(q_i -q_{i+1}), \ q_{i+N}=q_i
\end{equation}
The spectral curve equation for N-particle chain reads as
\begin{equation}
 y^2=P_N^2(x)-4\La^{2N}
\end{equation}
$P_N(x)$ is polynomial of degree N which encodes the values of integrals of motion. In the center-of-mass frame $\sum p_i =0$:
\beq
P_N(x) = x^N - u x^{N-2} + ...
\eeq
Coefficient $u$ is equal to the energy of the Toda chain.
Equivalent form of the spectral curve reads as:
\begin{equation}
 w+\cfrac{\La^N}{w}=P_N(x)
\end{equation}
We see that we have two punctures, $w=0$ and $w=\infty$. Correspondingly, we have two series of 2nd kind Abelian differentials $d\Om_i^+, d\Om_i^-$ 
and pertinent times $T_i^+, T_i^-$. However, it turns out that Whitham equations are consistent only if we restrict ourselves to the case $T_i^+=T_i^-$, that is,
we work with $d\Om_i=d\Om_i^+ + d\Om_i^-$. 

Seiberg-Witten meromorphic differential $dS_{SW}$ is given by:
\begin{equation}
 dS_{SW}=\cfrac{x P'_N(x) dx}{y(x)}=x \cfrac{dw}{w}
\end{equation}
It satisfies
\begin{equation}
\frac{\pr dS_{SW}}{\pr moduli} \approx holomorphic
\end{equation}
It is holomorphic apart from two second-order poles near $w=0$ and $w=\infty$.

Throughout the paper we will extensively use its periods:
\begin{eqnarray}
 a_i = \oint_{A_i} dS_{SW} \\ \nonumber
 a_i^D = \oint_{B_i} dS_{SW} \\ \nonumber
d \Om_1 = dS_{SW} - \sum_k \om_k a_k
\end{eqnarray}
and celebrated Seiberg-Witten prepotential $F(a)$:
\beq
\cfrac{\pr F(a)}{\pr a_i} = a_i^D
\eeq
It is useful to introduce vectors $\vec{U}^{(j)}$:
\begin{equation}
U^{(j)}_k=\cfrac{1}{2 \pi i}\oint_{B_k} d \Omega_j
\end{equation}
which obey the identity
\begin{equation}
U^{(1)} = \vec a_D - \tau \vec a.
\end{equation}

Very interesting observation, first made in \cite{Whitham95} is that 
\begin{equation}
 \cfrac{\pr dS_{SW}}{\pr \log \La} = d\Om_1
\end{equation}
Therefore, $dS$ can be chosen to be the Seiberg-Witten meromorphic
differential and $T_1=\log \La$  is the first Whitham time. It is possible to choose different normalizations
for the SW differential and Whitham times. In our case it is easy to show that:
\beq
\cfrac{\pr F}{\pr T_1} = 4 \pi i N u
\eeq

In what follows, we will often omit the $SW$ subscript.

The second crucial observation is that $dS$ coincides with the action differential $pdq$ for the Toda chain. Indeed, in case $N=2$,
the spectral curve reads as:
\beq
w+\cfrac{\La^4}{w}=x^2-u
\eeq
Change of variables $x=p, w=\La^2 \exp(q)$ leads to
\beq
2 \La^2 \cosh(q) = p^2 - u
\eeq
and $dS=pdq$

One can introduce several times
\begin{equation}
dS=\sum_{i=1}^{\infty} T_i d \hat \Om_i
\end{equation}
where $d \hat \Om_i $ obey the following requirements:
\begin{equation}
\frac{\pr d \hat \Om_i}{\pr moduli} \approx holomorphic
\end{equation}
and $\approx$ means that they have the same periods and behavior  near the punctures
\begin{equation}
d \hat \Om_i = (\xi^{-i-1}+O(1)) d \xi
\end{equation}

 It was argued in  \cite{marshakov07}  that higher
times correspond to the perturbation of the UV Lagrangian by single-trace $N=2$ vector superfield operators:
\begin{equation}
\label{eq:higher}
\mathcal{L}_T=\tau_0 \cfrac{1}{2}\int d^2 \theta d^2 \tilde \theta\ tr{\Phi^2} + \sum_{k>0} \cfrac{T_k}{k+1}
\int d^2 \theta d^2 \tilde \theta\ tr \Phi^{k+1}
\end{equation}
The first Whitham time $T_1$ is just a shift of UV coupling. In Appendix we will discuss the spectral curve
when higher times are switched on and derive generalized Bethe equations for this case, which hitherto has not been discussed in a literature.

\subsection{Whitham dynamics in the real case}
For completeness, let us recall the analogue of the
Whitham hierarchy for the case of the real phase space. It means that we consider a real dimension one curve on a two dimensional
real plane instead of a complex  curve.
Let us introduce
complex coordinates $\bar z, z$ then the curve is determined by
the equation
\beq
\bar z=S(z)
\eeq
We shall assume
that  $\bar z, z$  pair yields the phase space of some
dynamical system and  the curve itself corresponds to its
energy level. With this setup
it is clear that Poisson bracket between $\bar z$ and $z$
is fixed by the standard symplectic form:
\begin{equation}
 \{z,\bar z\}=1
\end{equation}

Let us remind the key points from \cite{mwz} where the Whitham hierarchy
for the plane curve was developed. The phase space interpretation
has been suggested in \cite{gorsky2000}.
The Schwarz function  $S(z)$ is assumed to be analytic in a domain including the curve.
Consider the map of the exterior of the curve
to the exterior of the unit disk
\beq
\omega (z)=\frac{z}{r} +\sum_{j} p_{j}z^{-j}
\eeq
where $\omega$ is defined on the unit circle. Introduce
the moments of the curve
\beq
t_n=\frac{1}{2\pi in} \oint z^{-n}S(z)dz, \ n < 0 
\eeq
\beq
t_0 = \frac{1}{2\pi i} \oint S(z)dz 
\eeq
\beq
v_n=\frac{1}{2\pi i} \oint z^nS(z)dz,\ n>0
\eeq
\beq
v_0=\oint log|z|dz
\eeq
which provide the following expansion for the Schwarz function
\beq
S(z)=\sum{k} t_k z^{k-1} +t_0 z^{-1} +\sum{k}v_k z^{-k-1}
\eeq
Let us define the generating function
\beq
S(z)=\partial_{z}\Omega(z)
\eeq
where
\beq
\Omega(z)=\sum_{k=1} t_k z^{k} +t_0 logz -\sum_{k=1}v_k z^{-k}k^{-1} - 1/2 v_0.
\eeq

One can derive the following relations
\beq
\partial_{t_0} \Omega(z)=\log \omega(z)
\eeq
\beq
\partial _{t_n} \Omega(z)=(z^n(\omega))_{+} +1/2 (z^n(\omega))_{0}
\eeq
\beq
\partial _{\bar{t_n}} \Omega(z)=(S^n(\omega))_{+} +1/2 (S^n(\omega))_{0}
\eeq
Therefore
we identify  $\log \om$ as angle variable and
the area inside the curve $t_0$ as the
action variable.
Let us denote by $(S(\omega))_{+}$ the truncated Laurent series with only
positive powers of $\omega$  kept and the $(S(\omega))_{0}$
is the constant term in the series.
The differential $d\Omega$
\beq
d\Omega = Sdz + \log\omega dt_0 +\sum (H_kdt_k -\bar {H_k} d \bar{t_k})
\eeq
yields the Hamiltonians and  $\Omega$ itself can be immediately
identified as the generating function for the canonical transformation
from the pair $(z,\bar {z})$ to the canonical pair $(t_0,\log \omega)$.

The dynamical equations read
\beq
\label{eq1}
\partial _{t_n} S(z)= \partial _{z}H_n(z)
\eeq
\beq
\label{eq2}
\partial _{\bar{t_n}} S(z) = \partial _{z}\bar {H_n}(z)
\eeq
and the consistency of (\ref{eq1}), (\ref{eq2}) yields the zero-curvature condition
which amounts to the equations of the dispersionless Toda lattice
hierarchy. The first equation of the hierarchy reads as follows
\beq
\partial^{2}_{t_1\bar {t_1}}\phi = \partial_{t_0}e^{\partial_{t_0}\phi}
\eeq
where $ \partial_{t_0}\phi =2\log r$.
The Lax operator L coincides with $z(\omega)$
\beq
L\Psi(z,t_0)=z\Psi
\eeq
and its eigenfunction -
Baker-Akhiezer(BA) function looks as follows
$\Psi=e^{\frac{\Omega}{h}}$.
Hamiltonians corresponding to  the Whitham dynamics
are expressed in terms of the Lax operator as follows
\beq
H_k=(L^k)_{+} +1/2(L^k)_{0}
\eeq

Now it is clear that the BA function is nothing but the coherent wave function
in the action representation. Indeed the coherent wave function
is the eigenfunction of the creation operator
\beq
\hat b \Psi=b\Psi
\eeq
From the equations above it is also clear that $\Omega$
it is the generating function for the canonical transformations
from the $b,b^{+}$ representation to the angle-action variables.

Having identified the BA function for the generic system let us
comment on the role of the $\tau$ function in the generic case. To this
aim it is convenient to use the following expression for the
$\tau$ function
\beq
\tau(t,W)=<t,\bar{t}|W>
\eeq
where the bra vector depends on times while the ket vector
is fixed by the  point of Grassmanian
\beq
|W>=S|0> , S=exp\sum_{nm}A_{nm}\bar{\psi}_{-n-1/2} \psi_{-m-1/2}
\eeq
This representation is convenient  for the
application of the fermionic language
\beq
\tau(t,W)=\frac{<N|\Psi(z_1)...\Psi(z_N)|W>}{\Delta(z)}
\eeq
where $\Delta(z)$ is Vandermonde determinant.

The consideration above suggests the following picture behind the
definition of the $\tau$ function. The fixing  the integrals of motion 
of the dynamical system  yields the curve on the phase space. Then
the domain inside the trajectory is filled by the coherent states
for this particular system. Since the coherent state occupies the
minimal cell of the phase space the number of the coherent states
packed inside the domain is finite and equals N. Since there is only one
coherent state per cell for the complete set it actually behaves
like a fermion implying a kind of the fermionic representation.

Therefore we can develop the second dynamical system of the Toda
type based on the generic dynamical system. The number
of the independent time variables in the Toda system  amounts from
the independent parameters in the potential in the initial
system plus additional time attributed to the action variable.
Let us emphasize that the choice of the particular
initial dynamical system amounts to the choice of the
particular solution to the Toda lattice hierarchy.


\section{Argyres-Douglas point in the Hamiltonian dynamics}
\label{sec:ad}
\subsection{Generalities}
Here we review the Argyres-Douglas phenomenon \cite{AD} and following \cite{newn2} demonstrate how one can compute some anomalous dimensions in the superconformal 
theory. The emergence of the conformal symmetry constitutes the AD phenomenon.

The key element of the Seiberg-Witten solution is the spectral curve which is $\qquad$ (N-1)-genus complex curve for $SU(N)$ gauge theory.
In case of pure gauge $SU(N)$ theory it is given by($\La$ is dynamical scale)
\begin{equation}
\label{eq:sc_Toda}
y^2=P(x)^2-\La^{2N}
\end{equation}
\begin{equation}
P(x)=x^N-\sum_{i=2}^N h_i x^{N-i}
\end{equation}
In $SU(2)$ case it is torus:
\begin{equation}
y^2=(x^2-u+\La^2)(x^2-u-\La^2)
\end{equation} 
where $u=h_2$, which at $u^2=\La^4$  degenerates - one of its cycles shrinks to zero.
Recalling BPS-mass formula, this can be interpreted as monopole/dyon becomes massless and the description of the low-energy theory
as U(1) gauge theory breaks down. Much more interesting situation is possible in $SU(3)$ case \cite{AD}:
\begin{equation}
P(x)=x^3-u x -v
\end{equation}
then for $u=0,v^2=\La^6$, the curve becomes singular:
\begin{equation}
y^2=x^3(x^3\pm2 \La^3)
\end{equation}
In this case, two \textit{intersecting} cycles shrink - it means that mutually non-local particles (monopole and dyon charged
with respect to the same U(1)) become
massless. In \cite{AD} it was conjectured that at this point the theory is superconformal. This result was
generalized to $SU(2)$ gauge theory with fundamental multiplets in \cite{newn2}.

In brief, the argument goes as follows:
Let us  denote
\begin{equation}
\label{eq:param}
\delta u=u=3 \ep^2 \rho,\ \delta v = v-\La^3 = 2 \ep^3
\end{equation}
$\rho$ is dimensionless, $\ep$ has a dimension of mass
and sets an energy scale.
Then the genus two curve degenerates to the "small" torus
\begin{equation}
\label{eq:smtor}
y^2=x^3-\delta u x -\delta v
\end{equation}
with modular parameter $\tau_{11}=\tau(\rho)+O(\delta v/\La^3)+O(\delta u/\La^2)$ and masses
\begin{equation}
\label{eq:as}
a^s, a^s_D \approx \epsilon^{5/2}/\Lambda^{3/2} \ra 0
\end{equation}
and periods $\om^s, \om^s_D \sim 1/a \ra \infty$.
The modular parameter of the  "large" torus $y^2=x(x^3 - \delta u x + \delta v + 2 \La^3)$  is  $\tau_l=\tau_{22}=e^{\pi i/3}+O(\delta u/\La^2)+O(\delta v/\La^3)$.
Below we will often use "s" and "l" indices to denote small and large tori.
The period matrix becomes diagonal (again up to $O(\delta u , \delta v)$ non-diagonal terms):
\begin{equation}
\tau=
\begin{pmatrix}
\tau(\rho)_s & 0 \\
0 & e^{\pi i /3}
\end{pmatrix}
\end{equation}

The crucial observation is that modulus of the "small" torus is independent of scale $\ep$. Due to the diagonal form
of the period matrix, the "small" U(1) factor (with masses $\approx \ep^{5/2}/\La^{3/2}$) decouples 
from the "large" U(1) factor (with masses $\approx \La$) and we are left with the RG fixed point with the coupling constant $\tau_s=e^{2 \pi i/3}$ - this
fact constitutes the Argyres-Douglas phenomenon.

Anomalous dimensions can be restored as follows \cite{newn2}. K\"ahler potential $\Im(a a_D)$ has dimension 2, so
$a$ and $a_D$ have dimension 1. From (\ref{eq:smtor}) we infer that relative dimensions are $D(x):D(\delta u):D(\delta v)=1:2:3$ -
it could be seen either as the $R-$charge condition or as a requirement for a cubic singularity. From (\ref{eq:as}) we see that
$D(\ep)=2/5$, therefore
\begin{eqnarray}
\label{eq:dims}
 D(x) = 2/5 \\ \nonumber
 D(\delta u)=4/5 \\ \nonumber
 D(\delta v)=6/5
\end{eqnarray}

\subsection{Toda chain: Argyres-Douglas point}
\label{sec:ad_toda}
In this subsection we comment on  the behavior of the solutions to the equations of motion of  Toda chain near the Argyres-Douglas point and show how 
the number of effective  degrees of freedom get reduced.

In the case of a periodic Toda chain  it is possible to write down an explicit solution using the so-called tau-function\cite{babelon}: 
\begin{equation}
\tau_n(t)=\theta(2 \pi i n \vec U^{(0)} + 2 \pi i t \vec U^{(1)} + \vec \zeta| \tau)
\end{equation}
$\zeta$ - is just a constant, $\vec U^{(k)}$ are defined in the section [\ref{sec:geom}]. Then  coordinates of particles $q_n$ can be expressed in terms of $\tau$-functions
\begin{equation}
\label{eq:solt}
\exp \l 2 \l q_{n}-q_{n+1} \r \r= \cfrac{\tau_{n+1} \tau_{n-1}}{\tau_n^2}
\end{equation}
Since at the AD point the period matrix is diagonal, the theta function factorizes into the product of two
theta functions corresponding to small torus and large torus:
\begin{equation}
\tau_n(t)=\theta(2 \pi i n  U^{(0)}_s + 2 \pi i t U^{(1)}_s + \zeta_s|\tau_s) \theta(2 \pi i n  U^{(0)}_l + 2 \pi i t U^{(1)}_l + \zeta_l|\tau_l)
\end{equation}
Moreover, since $\vec U^{(1)}=\vec a_D - \tau \vec a$ and 
$\tau_s=e^{2 \pi i /3}$, $a^s, a_D^s \ra 0$ corresponding theta function completely decouples and the solution is
determined up to the relative shift in terms of the large torus only. This is the reduction of degrees of freedom mentioned in the Introduction.

In principle, it is possible to carry out
more accurate analysis involving the effect from the non-diagonal terms of the period matrix. However, the only effect 
from these terms is a periodic modulation of the whole trajectory. After the averaging over large times these oscillations disappear.

In the case of N-particles it is possible to degenerate several pairs of intersecting cycles. In this case several small tori
will appear. The period matrix will be block-diagonal and respective masses $a, a_D$ tend to zero. So we can conclude that
small tori will again decouple and corresponding degrees of freedom get frozen.

\subsection{ Critical indexes in superconformal theory and Berry indexes}
Now we are going a propose a definition for "anomalies dimensions" and correction length for Toda chain near the AD point.
We will see that both mechanical and field-theoretical anomalous dimensions have the same nature as Berry indices in catastrophe theory.
 
Since the superconformal theory actually pertains to the small torus, lets look closely at the vicinity of the
Argyres-Douglas point. Near the AD point two tori are almost independent, so we can concentrate solely on the
part of the tau function which corresponds to the small torus - we will drop subscript $s$ for brevity.
The key observation above was that
$a,a_D \ra 0$ and $\tau \ra exp(2 \pi i/3)$, hence we can expand (\ref{eq:solt}) in Taylor series:
\begin{equation}
 \tau_n(t) \approx \tau_n(0)+ 2 \pi i B_n(a_D - \tau a) t
\end{equation}
We denote $\theta^\prime(2 \pi i n U^{(0)}+ \zeta| \tau) = B_n $ for brevity, then
\begin{equation}
\label{eq:prop}
2(q_n - q_{n+1}) =
\log \l \cfrac{\tau_{n+1}(0) \tau_{n-1}(0)}{\tau^2_n(0)} \r + 2 \pi i \l \cfrac{B_{n+1}}{\tau_{n+1}(0)} +
\cfrac{B_{n-1}}{\tau_{n-1}(0)} -2 \cfrac{B_n}{\tau_{n}(0)} \r (a_D - \tau a) t
\end{equation}

For general $\zeta$, the coefficient in front of $(a_D - \tau a) t$ is not zero. Let us recall  that the
modular parameter $\tau$ is independent of $\ep$ in the leading order. The same is true for the $U^{(0)}$ since it equals
to $L_D - L \tau$, where $L_D,L$ are periods of third kind Abelian differential $x^2 dx/y$
\begin{equation}
 \cfrac{x^2 dx}{y} \approx \cfrac{z dz}{\sqrt{z(z^3-3 \rho z -2)}},\ x=\ep z
\end{equation}
We can define "correlation length" $\delta q$ as the distance traveled by particles over the time $1/\Lambda$. Usually, correlation
length tends to infinity near a conformal point. Here, in classical mechanical system, it tends to zero. We can 
obtain "anomalous dimensions" by re-expressing the integrals of motion in terms of $\delta q$:
\begin{eqnarray}
\label{eq:an_def}
 \delta u = (\delta q)^\alpha,\ \delta v = (\delta q)^\beta
\end{eqnarray}
Equation (\ref{eq:prop}) tells us that $\delta q$ is proportional to $a$, that is, it has a field-theoretical anomalous dimension 1. Therefore, we have managed to
define mechanical "anomalous dimensions", given by (\ref{eq:an_def}), 
which coincide the field-theoretical anomalous dimensions (\ref{eq:dims}).

Surprisingly, counterparts of  these superconformal dimensions also arise in the context of caustics in optics
 (see \cite{berry} for a review). In optics,
one is interested in the wave function:
\begin{equation}
 \psi(\vec{C}) = \sqrt{k} \int \ ds \exp( i k W(s,\vec{C}))
\end{equation}
 where $k$ is an inverse wavelength and $W$ defines the geometry of light sources. One can define singularity indices $\beta, \sigma_j$ as
\begin{equation}
\label{eq:l_brane}
 \psi = k^\beta \Psi(k^{\sigma_j} C_j )
\end{equation}
(Note that $\Psi$ does not depend explicitly on $k$, so this definition is not meaningless).

Classification
of all possible $W$ has been intensively studied in the catastrophe theory framework. Equation (\ref{eq:l_brane}) reminds the wave
function of Lagrangian brane, with $W$ playing the role of the superpotential. From the SW theory viewpoint, $W$ defines the
spectral curve
\begin{equation}
 y^2=W(s,\vec{C})
\end{equation}
with $\vec{C}$ playing the role of moduli.  Let us consider the standard AD point in $SU(3)$. Then:
\begin{equation}
 y^2=(x^3-u x -v)^2-1=x^6 - 2 u x^4 - 2 v x^3 + ...
\end{equation}
In the notation of \cite{berry}:
\begin{equation}
 W(s)=s^6/6 + C_4 s^4 + C_3 s^3 + ...
\end{equation}
and singularity indices read as
\begin{equation}
 \sigma_4 = 1/3 \ \sigma_3 = 1/2
\end{equation}
However, we have to identify variables properly. In optics, or equivalently, classical mechanics  everything is measured in terms of $k$, whereas in the field theory everything is measured in terms of $a$.
Obviously, $x=s$ and $s$ has its own scaling properties: one requires the highest term $k s^6$ to be scale invariant \cite{berry}. Therefore $D(k)=6 D(s)=12/5$ (recall that $D(x)=2/5$ - eq. (\ref{eq:dims})).
Therefore, in the field theoretical normalization
\begin{equation}
 \sigma_4^{ft}=4/5,\ \sigma_3^{ft}=6/5
\end{equation}
 which are exactly the anomalous dimensions in the equation (\ref{eq:dims}). 
Similar analysis can be carried out for the case of $N_c=2, N_f=1$ where we have found a perfect agreement too. So we see that Berry indices have 
exactly the same nature as superconformal anomalies dimensions.

\subsection{Argyres-Douglas point via Whitham flows}
\label{sec:ad_wh}
In this section we specify the Whitham equations to the case of the pure $SU(3)$ gauge theory - 3 particle
Toda chain. Then we consider the $SU(2)$ case with fundamental matter. We demonstrate that the AD point, "small" and "big" tori
(in terminology of section \ref{sec:ad}) again decouple and the "small" torus is a fixed point for the Whitham dynamics.

First of all, let us make a comment about the maximum number of Whitham times we can introduce. Recall that 
the common wisdom of integrable systems dictates that we need exactly $N$ integrals of motion for a mechanical
system with $N$ degrees of freedom in order to the later be integrable.
From the point of view of N-particle closed Toda chain, higher times $T_l, l>N-1$ just do not exist and corresponding
flow should be trivial. From the field-theoretical viewpoint, it reflects the fact that for a $N\times N$ matrix A,
$A^N=a_1 A^{N-1}+..+a_{N-1} A + a_N$ - recall the interpretation of higher times via (\ref{eq:higher}). 
However, the trivialization of the flow from field-theoretical point of view is not  obvious. In order to prove that
\begin{equation}
\cfrac{\pr a_i^D}{\pr T_j} = 0,\  \operatorname{for} \  j>N-1
\end{equation}
we will use Riemann bilinear identity and equation (\ref{eq:whprep}).
In case of Toda-chain Abelian differentials of the first kind are linear combinations of $dx/y,...,x^{N-2} dx/y$,
therefore they have at most (N-2)-degree zero
at infinity, $d^{-1} d \Om_j$ has pole of order j at infinity. Taking $\tilde \om_1 = d \Om_j,\tilde \om_2 = \om_k$,
we obtain
\begin{equation}
\oint_{B_k} d \Om_j = 0 , j>N-1
\end{equation}
therefore the flow in indeed trivial.

In case of N-particle Toda chain all  first- and second- derivatives of prepotential were calculated in \cite{Whitham98}.
\begin{equation}
\cfrac{\pr F}{\pr T_n} = \cfrac{2 \pi i \beta}{n} \sum_m m T_m H_{m+1,n+1}
\end{equation}

\begin{equation}
\cfrac{\pr a^D_i}{\pr T_n} = \cfrac{2 \pi i \beta}{n} \cfrac{\pr H_{n+1}}{\pr a_i}
\end{equation}
\begin{equation}
\begin{split}
H_{m+1,n+1}=H_{n+1,m+1}=-\cfrac{N}{nm} res_{\infty} \l P(x)^{n/N} dP(x)_{+}^{m/N} \r \\
H_{m+1}=H_{m+1,2}= -\cfrac{N}{n} res_{\infty} \l P^{m/N}(x) dx \r
\end{split}
\end{equation}
where $(\sum_{n=-\infty}^{+\infty} a_n x^n)_{+}=\sum_{n=0} a_n x^n$. $\beta$ is one loop beta-function $\beta=2 N$. $P(x)$ defines
the Seiberg-Witten curve by $y^2=P(x)^2-\La^{2N}$. And in our normalization:
\beq
dS_{SW}=\cfrac{x dP}{y}
\eeq
This differential is $2 \pi i$ times greater than the one used in \cite{Whitham98}.

Since Argyres-Douglas point is RG fixed point for one of U(1) factors, we conjecture that for this U(1) factor (i.e. "small" torus)
Whitham dynamics should be also trivial at least in $T_1$. In case of $SU(3)$ we can consider only $T_1$ and $T_2$.
In this case  $H_{2}=u, H_{3}=v$ and using the fact that 
\beq
\cfrac{\pr dS_{SW}}{\pr H_i}=dv_{3-i}=\cfrac{x^{3-i} dx}{y}
\eeq
Whitham equations read as:
\begin{equation}
 \cfrac{\pr \vec{a}^D}{\pr T_1}=2 \pi i \beta \omega^{-1} \begin{pmatrix}0\\1\end{pmatrix}
\end{equation}
\begin{equation}
  \cfrac{\pr \vec{a}^D}{\pr T_2}=2\pi i \beta \omega^{-1} \begin{pmatrix}\frac{1}{2}\\0\end{pmatrix}
\end{equation}
where
\begin{equation}
 \omega_{kl}=\oint_{A_l} dv_k
\end{equation}
near Argyres-Douglas point \cite{AD,marino99},
\begin{equation}
 \omega=
\begin{pmatrix}
 - \cfrac{\ep^{-1/2} \om_\rho}{4 \pi \La^{3/2}} & \cfrac{d}{\La^2} \\
 2 \ep^{1/2} \eta & \cfrac{c}{\La}
\end{pmatrix}
\end{equation}
where $\omega_\rho$ is the period of the rescaled small torus - recall the section \ref{sec:ad}:
\beq
w^2=z^3-3 \rho z -2,\ x= \ep z,\ y=w \ep^{3/2}
\eeq
$\eta=\zeta(\om_\rho/2)$ is the value of Weierstrass zeta function at half-period, 
$c$ and $d$ are non-zero numerical constants. According to results of \cite{exp_cd}, they can be expressed as elliptic integrals:
\beq
\begin{split}
d=\cfrac{4}{i(r-1)^2} \int_0^1 d \xi \ \cfrac{1}{\sqrt{(1-\xi^2)(1-l^2 \xi^2)(\xi^2+k)^3}} \\
c=\cfrac{4}{i(r-1)} \int_0^1 d \xi \ \cfrac{1}{\sqrt{(1-\xi^2)(1-l^2 \xi^2)(\xi^2+k)}} \\
r=\exp(2 \pi i/3),\ l^2=-r,\ k=\cfrac{1}{r-1}
\end{split}
\eeq
Therefore,
\begin{equation}
 \omega^{-1}=
\begin{pmatrix}
 -\cfrac{4 \pi  \La^{3/2} \ep^{1/2}}{\om_\rho} & \cfrac{4 \pi \La^{1/2} \ep^{1/2} d}{\om_\rho c}  \\
\cfrac{8 \pi \eta  \La^{5/2} \ep}{\om_\rho  c} & \cfrac{\La}{c}
\end{pmatrix}
\end{equation}

So we conclude that all derivatives vanish, except
\begin{equation}
  \cfrac{\pr a^D_2}{\pr T_1}=12 \pi i \cfrac{\La}{c}
\end{equation}
It is not surprising since "large" torus is not degenerate and corresponding masses are $\approx \Lambda$. As we have promised before, Whitham
flow is stationary for the "small" torus, that is for the superconformal part of the theory.

We can rewrite the Whitham equations in a bit different form. In our case \cite{Whitham98}
\begin{equation}
 dS=T_1 dS_{SW} + T_2 d \hat \Om_2 = T_1 \cfrac{x(3x^2-u)dx}{y} + T_2 \cfrac{(3x^2-u)(x^2-2u/3)dx}{y}
\end{equation}
Applying Riemann bilinear identity to $dS$ and $dv_k$:
\begin{equation}
 \om^D \vec{a} - \om \vec{a}^D = 2 \pi i \begin{pmatrix} T_2/2\\T_1\end{pmatrix}
\end{equation}
and for $dv_k$ and $d \Om_1$($d \Om_i$ are defined in the section \ref{sec:geom}):
\begin{equation}
 \om \oint_{\vec{B}} d \Om_1 = 2 \pi i \begin{pmatrix}0\\1\end{pmatrix}
\end{equation}
\begin{equation}
 \om \oint_{\vec{B}} d \Om_2 = 2 \pi i \begin{pmatrix}\frac{1}{2}\\0\end{pmatrix}
\end{equation}
we can write:
\begin{equation}
\label{eq:w1}
 \cfrac{\pr \vec{a}^D}{\pr T_1} =
 \oint_{\vec{B}} d \Om_1 = \cfrac{1}{ T_1} \l \vec{a}_D - \tau \vec{a} - \cfrac{2 \pi i}{\omega} \begin{pmatrix}T_2/2\\0\end{pmatrix} \r
\end{equation}
\begin{equation}
\label{eq:w2}
 \cfrac{\pr \vec{a}^D}{\pr T_2} = \oint_{\vec{B}} d \Om_2 =
 \cfrac{1}{ T_2}  \l \vec{a}_D - \tau \vec{a} - \cfrac{2 \pi i}{\omega} \begin{pmatrix}0\\T_1\end{pmatrix} \r
\end{equation}
therefore 
\beq
T_1 \cfrac{\pr a_2^D}{\pr T_1}=a_2^D-e^{i \pi /3}a_2
\eeq
Generalization of (\ref{eq:w1}),(\ref{eq:w2}) for $SU(N)$ with non-zero times $T_i$ is straightforward.
This form clearly shows that if some U(1) factors decouple, they decouple in the Whitham dynamics as well. Whitham equations depend on the
choice of A- and B-cycles, in other words they are not invariant under modular group. AD point is significant because it is modular invariant. It 
means that whatever basis of cycles we choose, AD will be stationary point.

Let us compare the AD point with other possible degenerations, for example to the case when all B-cycles vanish \cite{douglas95}. For simplicity  take $T_2=0$ then the
period matrix:
\begin{equation}
 \tau_{mn} = -\cfrac{i}{2 \pi} \delta_{mn} \log{\cfrac{a^D_m}{\Lambda_m}}
\end{equation}
where $\Lambda_m$ - are some constants.
Due to the diagonal form of $\tau$, two U(1) factors again decouple.
Since $a_n^D \ra 0$ and $a_n$ do not vanish, Whitham dynamics is nontrivial.

If all A-cycles vanish,
\begin{equation}
 \tau_{mn} = -\cfrac{i}{2 \pi} \delta_{mn} \log{\cfrac{a_m}{\Lambda_m}}
\end{equation}
$a_n \ra 0 $, so dynamics is again nontrivial.

Now let us consider the $SU(2)$ theory with fundamental matter. General theory of Whitham hierarchy is a bit different 
in this case, because $dS_{SW}$ acquires additional poles, so we will not present the definition of the whole hierarchy.  
If $N_f<4$, beta-function is not zero and RG dynamics is not trivial. In \cite{eguchi}
the case with only two non-zero Whitham times was considered. The result is as follows: we have two non-zero times from the very beginning:
\begin{eqnarray}
 T_1  = \log(\La) \\ \nonumber
 T_0 = -\cfrac{1}{4 \pi i} \sum_{k=1}^{N_f} m_k
\end{eqnarray}
and the derivative of the prepotential with respect to $T_1$:
\begin{equation}
 \cfrac{\pr F}{\pr T_1} = 2 \pi i(2-\cfrac{N_f}{2})( 2 u - \cfrac{1}{2} \sum_{i=k}^{N_f} m_k^2)
\end{equation}
According to general philosophy, $\pr a/\pr T_1=0,\pr a/\pr T_0=0$ , hence for $N_f=1$
\begin{equation}
 \cfrac{\pr a^D}{\pr T_1} = 8 \pi i \cfrac{\pr u}{\pr a}
\end{equation}
We see that the right hand side is proportional to the charge condensate (see section [\ref{sec:mon}]). It was proved in  \cite{condensate}
that both monopole and charge condensate vanish at the AD point in the theory with $N_f=1$. Therefore, we contend that the statement that
\textit{the AD point is a fixed point for the Whitham dynamics} holds when fundamental matter is switched on.


\subsection{On confinement in the classical mechanics}
\label{sec:mon}
 Since the main purpose of the paper is to understand the reincarnation
 of the field theory phenomena in the complex classical dynamics we are to
 make some comment on the confinement phenomena. The rigorous derivation
 of the confinement in the softly broken N=2 SUSY YM theory in \cite{sw}
 was the first example  in the strongly coupled gauge theory. Although it is a kind
 of abelian confinement irrelevant for QCD  it is extremely interesting by its own.
 At the end of the subsection we will show that there is a very intimate new relation between the Konishi anomaly and
 Whitham equation.

 The non-vanishing order parameter  is the monopole condensate which provides
 the confinement of the electric degrees of freedom. It is proportional to the
 parameter of microscopic perturbation by $N=1$ superpotential 
\beq
\label{eq:ton1}
W_{UV}(\Phi)=\mu \tr \Phi^2 
\eeq
which breaks $N=2$ to $N=1$. In the IR one has the following exact superpotential \cite{sw}:
\beq
W_{IR}(\phi) = \mu u(\phi)
\eeq 
At the monopole point, where $a_D=0$, one arrives at the  monopole condensate \cite{sw}:
\begin{equation}
\label{monopole}
 <M \tilde M> = -\cfrac{\mu}{\sqrt{2}} \cfrac{\pr u}{\pr a^D}
\end{equation}
and the charge condensate of matter in the fundamental representation \cite{condensate}:
\begin{equation}
\label{charge}
 <Q \tilde Q> = -\sqrt{2}\mu \cfrac{\pr u}{\pr a}
\end{equation}

One more piece of intuition comes from the consideration of the AD point
in the softly broken SQCD \cite{condensate}. Since at the AD point
both monopole and matter condensates vanish  the AD point is
the point of deconfinement phase transition. Note that the gluino condensate
does not vanish at the AD point. These results have been obtained using the
interpolation between $N=2$ and $N=1$ theories via the Konishi anomalies.

We would like to ask a bit provocative question: is it possible to recognize
all condensates and the deconfinement phase transition in the framework of the
classical mechanics? We shall not answer these questions completely but make some
preliminary discussion on this issue. First of all, consider the pure $SU(2)$  case
which corresponds to the cosine potential. Upon the perturbation added the
monopole condensate (\ref{monopole})  gets developed and due to the Konishi
anomaly relation the gluino condensate is proportional to the scalar condensate
\beq
\label{eq:konishi}
<\lambda \lambda> =  -8 \pi^2 \mu <\tr \phi^2> = -4 \pi^2  <\phi \cfrac{\pr W_{UV}}{\pr \phi}>
\eeq
Therefore, as the first step we could ask about the meaning of the Konishi anomaly
relation in the Hamiltonian framework. Two dynamical systems are involved. The scalar
condensate $u$ plays the role of the energy in the $N=2$ Hamiltonian system
with $V=\Lambda \cos q$ while upon deformation to $N=1$, the gluino condensate plays the role of the action(period of 
1-point resolvent) in the Dijkgraaf-Vafa matrix model \cite{dv0206,dv0207}. Potential for this system reads as:
\beq
V=W_{UV}'^2 + f_{n-1}
\eeq
where $f_{n-1}$ is polynomial of degree $n-1$, if $W_{UV}$ has degree $n+1$.
For the simplest deformation $\mu \Phi^2$ it is nothing but the complex oscillator.

Actually we have to make the second step. At the first one the meaning of the AD
point as the decoupling of the small torus has been found. Now the question
concerns the very precise identification of the soft breaking of SUSY in the
framework of the complex Hamiltonian system. The analogy with the Peierls model
mentioned in \cite{peierls} can be useful here. It describes the one-dimensional
superconductivity of electrons propagating on the lattice. The key point is that the
Riemann surface which is the solution to the equation of motion in the Toda system
simultaneously plays the role of the dispersion law for the Lax fermions. Therefore
the degeneration of the surface at AD point corresponds to the degeneration  of the Fermi surface
for the fermions. Therefore the deconfinement phase transition at AD points presumably
corresponds to the breakdown of superconductivity in the Peierls model.
We hope to discuss this issue in details
elsewhere.

Also, note that eqs. (\ref{monopole}),(\ref{charge}) strongly resemble Whitham equations of motion from the 
previous section. It is not a coincidence - Whitham dynamics is useful for softly breaking $N=2 \ra N=0$ \cite{marino1,marino2,marino3}:
we can promote the first time $T_1=\log \La$ to background $N=1$ spurion chiral multiplet. After that, we can switch on the other scalar component of this
multiplet:
\begin{equation}
 T_1=\log \La + \theta^2 G
\end{equation}
This deformation preserves all holomorphic properties of the original theory, so we are able to write down \textit{the exact}
 prepotential for this new theory:
\begin{equation}
 \tilde F=F(G=0)+\cfrac{\pr F}{\pr T_1} G \theta^2
\end{equation}
Since $\theta$ explicitly enters the prepotential, the theory has no supersymmetry. Additional terms in the IR Largangian are\cite{marino1}($G^*=G$):
\begin{equation}
\label{eq:ir_f}
 \Delta \mathcal{L}_{IR}=\cfrac{1}{8 \pi}(\la \la + \psi \psi) \Im(\cfrac{\pr F''}{\pr T_1})G + \cfrac{1}{4 \pi \tau } 
 \Im(\bar \phi \cfrac{\pr F''}{\pr T_1}) \Im(\cfrac{\pr F'}{\pr T_1})G^2
\end{equation}
where $F'=\pr F/\pr a$ and $\tau=\Im(F'')$ - is a coupling constant, $\psi$ is a fermion in the $N=1$ chiral multiplet. In the UV we have:
\begin{equation}
 \Delta \mathcal{L}_{UV}=(\la \la + \psi \psi) G + \cfrac{1}{\tau} \Im(\phi)^2 G^2
\end{equation}
Note that $G$ gives masses to both fermions and imaginary part of the Higgs field, whereas deformation to $N=1$ by the superpotential (\ref{eq:ton1})
gives usual Higgs mass term $\mu^2 \bar \phi \phi$ and $\mu \psi \psi$ and does not give mass to the gluino $\lambda$. In \cite{marino2,marino3} various
monopole and dyon condensates were calculated. Here, we find gluino condensate, that is we derive an analogue of the Konishi anomaly using Whitham equations. 
Let us emphasize once more that we deal with not $N=1$ theory, but with the $N=0$ one obtained by a very special deformation of the $N=2$ theory. So we do not 
expect that the final expression would be the same as in  the $N=1$ theory. However, as we will see in a moment, the result naturally generalizes 
the Konishi anomaly.

Varying (\ref{eq:ir_f}) with respect to $\phi$ and $\la \la, \psi \psi$ (for simplicity we consider real $\phi$) we get:
\begin{equation} 
<\Im(\cfrac{\pr F''}{\pr T_1})> =0
\end{equation}
and taking into account that  $\pr F/\pr T_1=2u$
\begin{equation}
 <\psi \psi> + <\lambda \lambda> = -\cfrac{2}{\tau} <\phi> <\Im(\cfrac{\pr F'}{\pr T_1} )> = -\cfrac{4}{\tau} <\phi> <\Im \l \cfrac{\pr u}{\pr \phi} \r>
\end{equation}
Since $W_{IR}=\mu u$, the last equation looks very natural  and to some extend is an analogue of  (\ref{eq:konishi}).



\section{On the quantization procedure }
\subsection{Different quantizations of  complex Hamiltonian systems}
\label{sec:diff_quant}
Here we review recent developments in quantization of complexified Hamiltonians systems. After that, we will demonstrate that
the curve of marginal stability(CMS) in the Seiberg-Witten theory is exactly the place where the level-crossing in such systems occur. To the
best of our knowledge, this interpretation of the CMS has never been proposed yet. 

There are some new points in the quantization of complex integrable systems. First of all, the essential part
of a quantization concerns a choice of Hilbert space. In the pioneer work \cite{bender98}, in the case of one degree of freedom 
the following quantization was suggested: Hilbert space consists of analytic functions on a complex plane with possible irregular singularity at infinity, and a scalar product is given by:
\begin{equation}
<\psi|\phi> = \int_\mathcal{C} \psi^*(q) \phi(q) dq
\end{equation}
where $\mathcal{C}$ is some contour on a complex plane. Hamiltonian is taken to be a standard one: $\hat H=\hat p^2/2 + U(q)$, with
$\hat p = i \pr/\pr q$. Then the Schr\"odinger equation
\begin{equation}
\hat H \psi = - \cfrac{\psi^{\prime \prime}(q)}{2} + U(q) \psi = E \psi(q)
\end{equation}
is just the standard Schr\"odinger equation analytically continued to a complex plane. If $U(q)$ is an entire function
then the equation is consistent with the definition of the Hilbert space. When the curve $\mathcal{C}$ coincides
with the real axis this construction gives the standard quantization.

In the real case the quantization condition for the energy levels comes from the requirement that the wave function is
normalizable. In \cite{bender98} 
an analogue of the WKB quantization was suggested:
\begin{equation}
\label{eq:ben_q}
 a(u)=\oint \sqrt{2(u-U(q))} dq= 2 \pi \hbar n, \ n \in \mathbb{N}
\end{equation}
where integral should be taken along the line where integrand is real.
Note that since everything is complex now, it is actually \textit{two} real conditions on a complex energy $u$:
\begin{eqnarray}
 \Re a = 2 \pi \hbar n \\ \nonumber
 \Im a = 0
\end{eqnarray}
Perfect agreement with numerical computations has been found. It worths mentioning that the same condition was proposed in \cite{kam_korot}
for studying complex non-hermitian Hamiltonians.

However, if the potential is not holomorphic, one can impose  different quantization condition:
wave function is not required to be holomorphic. Instead, one imposes its \textit{single-valuedness}. At least
one such example is known in literature \cite{gor_kor}: spectrum of XXX chain with \textit{complex spin}
emerging in high energy QCD for describing effective interaction between Reggeons \cite{kor01,kor02}. In brief, the problem
is as follows: complex spin chain has a \textit{non-holomorphic}
Hamiltonian:
\begin{equation}
 \mathcal{H}_{N} = H_N(z)^{s=0}+\bar H_N(\bar z)^{s=1}
\end{equation}
Actually $z$ and $\bar z$ are complex coordinates
on a \textit{real} plane of Reggion coordinates. 
Requirement that the $\psi$ has no monodromy around cycles yields a bit different WKB quantization condition \cite{gor_kor}:
\begin{eqnarray} 
\label{gkk_cond}
 \Re a = 2 \pi \hbar n \\ \nonumber
 \Re a_D = 2 \pi \hbar n_D
\end{eqnarray}
which coincides with the conventional WKB condition when $n_D=0$.

Returning to the SW theory, in \cite{nek_shat} it was shown that in the Nekrasov-Shatashvili(NS) limit $\ep_2=0$ of $\Omega$ deformation, 
underlying integrable systems get quantized.
The following quantization condition was proposed for theories without matter (Toda chain) or with adjoint matter (Calogero system):
\begin{equation}
\label{eq:ns}
 a_l = 2 \pi \ep_1 n_l
\end{equation}
Quantization condition (\ref{eq:ben_q}) looks exactly the same as Nekrasov-Shatashvili quantization. Nevertheless they are different:
in (\ref{eq:ben_q}) the integral can be taken along the finite number of paths on a complex plane(to ensure convergence), whereas in Nekrasov-Shatashvili quantization (\ref{eq:ns}) one can
choose arbitrary element of $SL(2, \mathbb{Z})$: the choice $a_l = 2 \pi \ep_1 n_l$ is called type A quantization condition, while $a_D = 2 \pi \ep_1 n_l$ - type B.
It was conjectured \cite{nek_shat} that the type A condition fixes the wave function to be normalizable on the \textit{real axis and} type B corresponds
to the wave function, which is $2 \pi$ periodic along the \textit{imaginary axis}.The  conjecture about the type A was proven in \cite{tba_toda}. We do not
know what conditions are imposed on the wave function by other elements of $SL(2, \mathbb{Z})$.

The case with fundamental matter was considered in \cite{bethe_xxx}, where it was shown that the conventional algebraic Bethe
ansatz with polynomial Baxter function implies $a_l=m_l-\ep_1 n_l,\ n_l \in \mathbb{N} $. In the Appendix we will show how this quantization
condition is modified by the non-zero Whitham times.


It is in order to make a comment concerning the place of  the curve of marginal stability in the quantum spectrum. In the Seiberg-Witten theory with the gauge group $SU(2)$ a BPS particle with
electric and magnetic charges $(q,p)$ has mass $M=Z=|q a + p a_D|$. A BPS particle can decay into a BPS particle iff $a$ and $a_D$
are collinear, that is
\begin{equation}
 \Im \cfrac{a_D}{a} = 0
\end{equation}
This equation defines the curve of marginal stability on the moduli space. 

On a quantum mechanical side, energy level crossing occurs when there are two different cycles with
the same allowed energy level. Let us denote these cycles $a$ and $n a + m a_D$. Nekrasov-Shatashvili quantization conditions:
\begin{equation}
a=k_1 \hbar
\end{equation}
\begin{equation}
 n a+ m a_D = k_2 \hbar
\end{equation}
$k_1,k_2 \in \mathbb{N}$, but $\hbar=\epsilon_1$ is not necessary real. If we divide the second equation by the first one
\begin{equation}
 m \cfrac{a_D}{a} = \cfrac{k_2}{k_1} - n
\end{equation}
If the original cycles are different, $m \neq 0$ and  $\Im{a_D/a}=0$. So we conclude that the level crossing can happen on the curve
of marginal stability only.


\subsection{Quantization and the Dunne--\"Unsal relation}
\label{sec:du}
In this section we investigate how Whitham equations are deformed by the Omega-deformation. We derive their explicit form for Toda chain in
 general Omega-deformation. Then, we will consider quantum mechanical particle in double-well potential and derive Whitham equations for this system.
We will use our results to show that Dunne-\"Unsal(D\"U) relation coincides with Whitham equations at least in the first order in Plank constant. 
This is one of our main results.

Seiberg-Witten solution to the Whitham-Krichever hierarchy can be thought of as a non-autonomous Hamiltonian system with the Hamiltonian 
$4  \pi i N u(a,\La)$ and
canonical pair $\{a^j,a^k_D\}=\delta^{j k}$ \cite{lns}. For 2-particle Toda chain:
\begin{eqnarray}
\label{eq:whiteq}
\cfrac{\pr a}{\pr \log{\La}} = 8 \pi i \cfrac{\pr u(a,\La)}{\pr a_D} = 0 \\ \nonumber
\cfrac{\pr a_D}{\pr \log{\La}} =  a_D - \tau a = \cfrac{8 \pi i}{\om} = 8 \pi i  \cfrac{\pr u(a, \La)}{\pr a}
\end{eqnarray}
The last equation follows from the Matone relation \cite{matone}:
\begin{equation}
2F - a a_D = \cfrac{\pr F}{\pr \log{\La}} = 8 \pi i u
\end{equation}
which, in turn, can be thought of as a Hamilton-Jacobi equation, where the prepotential is playing the role of the mechanical action.

In what follows we  will need to know  how Whitham dynamics is affected by the $\Omega$ deformation.
The prepotential involves two contributions \cite{nek2002}:
\begin{equation}
F_{Nek} = F_{inst}+F_{pert}
\end{equation}

\begin{equation}
F_{inst} = \sum_n q^{2 N n} \mathcal{F}_n,\ q=\cfrac{\La}{a}
\end{equation}
and  it was shown  in \cite{pog}   that the $\log \La$ derivative of the instanton part is unchanged by the $\Omega$-deformation:
\begin{equation}
u=\sum_k \l \cfrac{a_k}{2 \pi i} \r^2 + \cfrac{1}{2 \pi i}\sum_n n q^{2Nn} \mathcal{F}_n = \sum_k \l \cfrac{a_k}{2 \pi i} \r^2 
+ \cfrac{1}{4 \pi i N}\cfrac{\pr F_{inst}}{\pr \log{\La}}
\end{equation}
Factors $2\pi i $ appear because we adopted a bit different normalization for the SW differential.
\begin{equation}
F_{pert} = \ep_1 \ep_2 \sum_{l \neq n} \int_0^{+\infty} \cfrac{ds}{s} \cfrac{\exp(-s(a_l-a_n)/2\pi i)}{\sinh(s \ep_1/2) \sinh(s \ep_2/2)}
\end{equation}
The integral is divergent at the lower bound. The prescription is that one should keep only non-singular part - this is the origin
of the scale $\La$. Proper coefficient can be found by comparison with the known 1-loop expression. 
Expanding the integrand near $s = 0$, one obtains the following  $\La$-dependent terms:
\begin{equation}
4 \pi i N \sum_n \l \cfrac{a_n}{2 \pi i} \r^2 \log(\cfrac{a_n}{2 \pi i \La}) - 4 \pi i N \cfrac{\ep_1^2 + \ep_2^2}{24} \log{\cfrac{a_n}{2 \pi i \La}}
\end{equation}

Combining together perturbative and instanton contributions:
\begin{equation}
\cfrac{\pr F_{Nek}}{\pr \log{\La}} = 4 \pi i N \l u-\cfrac{\ep_1^2+\ep_2^2}{24} \r
\end{equation}
Upon differentiating w.r.t. $a$, we conclude that Whitham equations of motion (\ref{eq:whiteq}) still hold even in the case
of general $\ep_1,\ep_2$.

The natural question is what happens with the full Whitham hierarchy (\ref{eq:whbig}). One can try to attack this problem
 using beta-ensemble approach \cite{beta_fin_n,agt_beta}. This approach is based on the AGT conjecture, since conformal 
blocks are equal to Dotsenko-Fateev beta-ensemble(matrix model with deformed measure) partition function with \textit{finite N}
\cite{df}.  Actually, AGT conjecture in the NS limit($\ep_2 \ra 0$ which implies $N \ra \infty$ in the beta-ensemble) is equivalent to
the  following proposal of \cite{wkb_nek}:
WKB approximation allows one to expand the phase of the  wave function in powers of $\hbar=\ep_1$
\begin{equation}
\label{eq:wkb_ds}
\psi_{exact}(x) = \exp \l \cfrac{i}{\hbar}\int^x p_{quant} dq \r = \exp \l \cfrac{i}{\hbar} \l \int^x p dq + O(\hbar) \r \r
\end{equation}
In \cite{wkb_nek} it was conjectured that the prepotential obtained by computing WKB quantum periods
\beq
\begin{split}
a_{WKB} = \oint_A p_{quant} dq,\ a^D_{WKB} = \oint_B p_{quant} dq \\
a^D_{WKB}=\cfrac{\pr F^{WKB}}{\pr a}
\end{split}
\eeq
coincides with
the Nekrasov prepotential in the NS limit.
This statement was checked \cite{wkb_nek,popolit} up to $o(\hbar^6,\log \La)$ however no conceptual proof is known so far. At the end of
this section we will return to this conjecture.

On the other hand, in \cite{chekhov} the large N limit of the beta-ensemble was
thoroughly considered, and it was proven that the large N limit corresponds to the quantization of some mechanical system.
One point resolvent plays the role of the Seiberg-Witten meromorphic differential, moreover it equals to $d \psi/\psi$, where $\psi$ is wave-function of the quantum mechanical system.
We see that the AGT conjecture, the beta-ensemble approach and the conjecture about the exact WKB periods are all tightly related.
Strikingly, after an appropriate
deformation of Abelian meromorphic differentials, equations $(\ref{eq:whprep})$ and (\ref{eq:whprep}) still hold \cite{chekhov}.
Therefore if we believe in either the conjecture about the exact WKB periods (\ref{eq:wkb_ds}) from \cite{wkb_nek} or the AGT conjecture \cite{agt}, we can conclude that \textit{in the
Nekrasov-Shatashvili limit the Whitham dynamics is not quantized but only deformed.}

Moreover, using this conjecture we will show now that the Whitham equations in the form (\ref{eq:whiteq}) are quite general and are not affected by the quantization. For 
simplicity we will concentrate on genus one case. Let us consider Hamiltonian
\begin{equation}
H=\cfrac{p^2}{2} + c V(q) 
\end{equation}
$V(q)$ is polynomial of degree $2d$, $d>1$. For the exact WKB phase $p_{quant}=f$ we have the Riccati equation:
\begin{equation}
-i \hbar f' + f^2=2(E-cV(q))
\end{equation}
$f$ has a representation in power series in $\hbar$: $f=f_0 + \hbar f_1 + \hbar^2 f_2+...$
Several first terms are:
\begin{eqnarray}
f_0 = \sqrt{2(E-cV)} \\ \nonumber
f_1=-i \cfrac{cV'}{4(E-cV)} \\ \nonumber
f_2= \cfrac{1}{32} \cfrac{5c^2 V'^2 + 4cV''(E-cV)}{\sqrt{2} (E-cV)^{5/2}}
\end{eqnarray}
Again, since we require $\pr a/\pr c =0$, we have 
\begin{equation}
\cfrac{\pr a}{\pr c} = \oint_A \cfrac{\pr f}{\pr c} dq = 0
\end{equation}
Now we apply Riemann bilinear identities for differentials $\cfrac{\pr f}{\pr c} dq$ and $\cfrac{\pr f}{\pr E} dq$:
\begin{equation}
\cfrac{\pr a^D}{\pr c} \cfrac{\pr a}{\pr E} = 
\oint_A \cfrac{\pr f}{\pr E} dq \ \oint_B \cfrac{\pr f}{\pr c} dq = 2 \pi i \res_\infty \l \cfrac{\pr f}{\pr c}dq \  d^{-1} \l  \cfrac{\pr f}{\pr E} dq \r \r
\end{equation}
At the first sight, we have to add contributions from turning points where $E=cV$ and so $f_n, n>1$ have poles. However, these poles are artifacts of WKB 
method and exact wave function does not have any singularities apart from the one at infinity. Therefore, we do not have to take them into account.

The idea is that only $f_0$ contributes to the residue. Indeed, it is not difficult to show that at infinity: 
\beq
f_k = O(x^{-(1+(k-1)(d+1))}),\ x \ra \infty
\eeq
 and
\beq
\cfrac{\pr f_n}{\pr E} = O(x^{-(1+2d+(n-1)(d-1))}),\ x \ra \infty
\eeq
We conclude the contribution of order $\hbar^{n+k}$ is given by a differential which behaves at most as 
$O \l \cfrac{1}{x^{ 1+2d+(n+k-2)(d+1)}} \r$. 
The "classical" part $n=k=0$ behaves as $O(x)$ and 
therefore can have a non-trivial contribution, whereas quantum corrections are suppressed by powers of $x$. 
The first quantum correction, $n+k=1$, behaves as $O(1/x^d)$ so has a zero residue. Higher
quantum corrections have a zero even of higher degree at infinity. So we conclude that 
\begin{equation}
\cfrac{\pr a^D}{\pr c} = const \cfrac{\pr E}{\pr a}
\end{equation}
and $const$ depends on a normalization and does not receive quantum corrections.

Recently, there was  much progress in studying the relation between perturbative and non-perturbative expansions
(see \cite{zinn04,dunne12,dunne13, dunne14} and references therein) in both quantum mechanics and quantum field theory.
In \cite{zinn04} Zinn-Justin and Jentschura  using resurgence in multi-instanton expansion have  conjectured \textit{the exact}
quantization condition for several quantum mechanics potentials.
Amazingly it involves only two functions $B(E,g)$ and $A(E,g)$, where $E$ is an energy($u$ in our notation) and $g$ is a coupling constant.
In \cite{dunne13} Dunne and \"Unsal have found a relation between these two functions. We shall demonstrate that
this relation is nothing but Whitham equation of motion.

The most simple example is a double-well potential:
\begin{equation}
\label{eq:dwell}
H =\cfrac{p^2}{2} + \cfrac{1}{2} q^2(1- \sqrt{g}q)^2
\end{equation}
The first Whitham time is the coupling constant $c$ which stands in front of the whole potential $c V(q)$. In case of the double-well potential (\ref{eq:dwell})  $c$ coincides 
with $1/g$ and the rescaling $E \ra 2E/g$ is needed. In genus one, we have usual definitions for periods:
\begin{eqnarray}
a = \oint_A p dq \\ \nonumber
\om = \cfrac{\pr a}{\pr E}= \oint_A \cfrac{dq}{g \sqrt{2E/g-V(q)}}
\end{eqnarray}
The electric period $a$ corresponds to classically allowed region near the bottom of the well, whereas $a^D$ is
an instanton factor corresponding to the barrier penetration between two wells.

Let us recover coefficients in Whitham equations. If we  impose the constraint $\pr a/\pr g=0$ then we have for the dual period:
\begin{equation}
\cfrac{\pr a^D}{\pr g} = \cfrac{1}{g} \l \cfrac{\om^D}{\om} a - a^D \r
\end{equation}
Taking into account the Picard-Fuchs relation:
\begin{equation}
a^D \om - a \om^D = \cfrac{2 \pi i}{3}
\end{equation}
we get
\begin{equation}
\label{eq:dw_wh}
g^2 \cfrac{\pr a^D}{\pr g} =\cfrac{2 \pi i}{3} \cfrac{\pr E}{\pr a}
\end{equation}
and exact quantization condition reads as \cite{zinn04}(from now on we put $\hbar=1$, $\pm$ on the RHS distinguishes odd and even energy levels):
\begin{equation}
\label{eq:zj_dw} 
\cfrac{1}{\sqrt{2 \pi}} \Gamma \l \cfrac{1}{2}-B \r \l - \cfrac{2}{g} \r^B \exp(-A/2) = \pm i
\end{equation}
One should understand this relation in a sense that after finding the energy in series of $g$ (including non-perturbative factors) it will
be possible to resum the resulting series using Borel method. Moreover, all the ambiguities will cancel each other\cite{zinn04}. 

The Dunne-\"Unsal relation \cite{dunne13} reads as
\begin{equation}
\label{eq:du_dw}
\cfrac{\pr E(B,g)}{\pr B}= -6Bg-3g^2 \cfrac{\pr A(B,g)}{\pr g}
\end{equation}
where the function $B(E,g)$ is  easy to calculate 
\begin{equation}
B=\cfrac{a}{2 \pi}=\cfrac{1}{2 \pi} \oint_A p_{quant }dq = \cfrac{1}{2 \pi} \oint_A \sqrt{2E/g-V(q)} dq + O(\hbar)
\end{equation}
Originally,  calculation of the function $A(E,g)$ involved tedious multi-instanton calculation.
Note the arguments of $A(B,g)$: derivative w.r.t. $g$ is taken keeping $B$ constant. Since $B=a/2\pi$ we discern here the first Whitham equation $\pr a/\pr g=0$.
The second equation turns out to be  the Dunne-\"Unsal relation itself. Let us compare (\ref{eq:zj_dw}) with WKB quantization condition
for a double-well potential\cite{jwkb}(again, $\pm$ accounts for even and odd wave-functions):
\begin{equation}
\label{eq:q_dw}
\pm 1=\cfrac{1}{2} \exp(-ia^D/2) \cfrac{\sin(a/2)}{\cos(a/2)}
\end{equation}
The technical subtlety why we can not extend our claim about the connection between D\"U relation and Whitham equations is that (\ref{eq:q_dw})
is true only in the first order in Plank constant since its derivation uses quadratic approximation near the turning points.

From this we infer that
\begin{equation}
\log \l  \exp(-ia^D/2) \cfrac{\sin(a/2)}{\cos(a/2)} \r = \log \l \const \Gamma \l \cfrac{1}{2}-B \r \l - \cfrac{2}{g} \r^B \exp(-A/2) \r
\end{equation}
and taking the derivative w.r.t. $g$ at constant  $a$  we get
\begin{equation}
\cfrac{\pr A(B,g)}{\pr g} = i \cfrac{\pr a_D}{\pr g} - \cfrac{2 B}{g}
\end{equation}
 \begin{equation}
2 \pi \cfrac{\pr E }{\pr a} =-3 i g^2 \cfrac{\pr a^D}{\pr g}
\end{equation}
which is exactly the second Whitham equation of motion (\ref{eq:dw_wh}).

Another example is the sine-Gordon potential
\begin{equation}
E=\cfrac{p^2}{2}+\cfrac{1}{8} \sin(2 \sqrt{g} q)
\end{equation}
Identification between $E,g$ and usual parameters in Toda chain $u,\La$ reads as:
\beq
\label{eq:change}
u=-\cfrac{E}{2g},\ 2 \La^2=\cfrac{i}{16g}
\eeq
The Dunne--\"Unsal relation in this case reads as follows
\begin{equation}
\label{eq:du}
\cfrac{\pr E(B,g)}{\pr B}= -2Bg-g^2 \cfrac{\pr A(B,g)}{\pr g}
\end{equation}
According to \cite{zinn04}, exact quantization condition reads as
\begin{equation}
\label{eq:zj_sin}
\l \cfrac{2}{g} \r^{-B} \cfrac{\exp(A/2)}{\Gamma(1/2 - B)} + \l-\cfrac{2}{g} \r^B \cfrac{\exp(-A/2)}{\Gamma(1/2+B)} = \cfrac{2 \cos(\phi)}{\sqrt{2 \pi}}
\end{equation}
where $\phi$ is Bloch phase - we are dealing with the periodic potential which possesses band structure. 
Note the mismatch in the factor $1/2$ with the quantization condition obtained in \cite{dunne13}  using uniform WKB method \cite{dunne14} instead of
resurgence in instanton calculus. We argue
that the right choice is
\begin{equation}
\l -\cfrac{2}{g} \r^B \ra \l \cfrac{2}{g} \r^B  \cfrac{\cos(\pi B)}{2}
\end{equation}
We will show in a moment, that this analytical continuation  agrees with the Whitham equations, like in the double-well case.

To this end we can make use of the WKB quantization condition for a generic periodic potential, which can be obtained 
along the same lines as \ref{eq:q_dw}:
\begin{equation}
 2 \exp(i a^D/2) \cos(a/2) + \cfrac{1}{2} \exp(-i a^D/2) \cos(a/2) = 2 \cos(\phi)
\end{equation}
where $a$ and $a^D$ are electric and magnetic quantum periods as before.
Since
\begin{equation}
\cfrac{1}{\Gamma(1/2-B) \Gamma(1/2+B)} = \cfrac{\cos(\pi B)}{\pi}
\end{equation}
there is a very simple relation
\begin{equation}
\label{eq:simple}
2 \exp(i a^D/2) \cos(a/2) = \sqrt{2 \pi} \l \cfrac{2}{g} \r^{-B} \exp(A/2) \cfrac{1}{\Gamma(1/2-B)}
\end{equation}
which yields
\begin{equation}
\label{eq:compar}
\cfrac{\pr A(B,g)}{\pr g} = i \cfrac{\pr a_D}{\pr g} - \cfrac{2 B}{g}
\end{equation}
Substituting the  equation above into the Dunne-\"Unsal relation (\ref{eq:du}) we arrive at
\begin{equation}
2 \pi \cfrac{\pr E }{\pr a} =-i g^2 \cfrac{\pr a^D}{\pr g}
\end{equation}
Taking into account the change of variables (\ref{eq:change}) we obtain exactly the Whitham equations of motion (\ref{eq:whiteq}).

We would like to emphasize that we have derived Whitham equations including all quantum corrections, whereas
we have justified the connection between D\"U relation and Whitham equations only in the first order in Plank constant. The problem is that in
the WKB expansion it is not clear how to take into account transitions near turning points beyond the first two orders in Plank constant.

Fortunately, in case when the potential has strictly one non-degenerate minimum, in other words, \textit{only two simple turning points},
it is possible to obtain an exact WKB quantization condition\cite{one_well,voros}. In fact, for $V(q)=2 \La^2 \cosh(q)$,
it exactly coincides with the NS quantization condition:
\beq
\label{eq:onshell}
\oint_A p_{quant} dq =  a_{WKB}(u) = a_{Nek}(u) = a = 2 \pi n, n \in \mathbb{N}
\eeq
Also, as we have found: 
\beq
\label{eq:der}
\cfrac{\pr F^{WKB}}{\pr \log \La} = \cfrac{\pr F^{Nek}}{\pr \log \La} = 8 \pi i \l u-\cfrac{1}{24} \r
\eeq
(by $F^{WKB}$ we understand the prepotential obtained via the exact WKB periods)

Equation (\ref{eq:onshell}) holds only "on-shell", whereas equation (\ref{eq:der}) is true for any value of energy. Therefore, 
unfortunately, we can not prove rigorously that $F^{WKB}=F^{Nek}$. However, basing on above equations and on an explicit 
calculations made in \cite{wkb_nek,popolit}, we will assume that $F^{Nek}=F^{WKB}$. In \cite{nek_shat} it was argued that after the S-duality,
the NS quantization (\ref{eq:onshell}) leads to the condition of $2\pi$-periodicity of the Bloch-wave in the potential $2 \La^2 \sin(q)$:
\beq
\label{eq:s_ns}
a^D = \cfrac{\pr F^{Nek}}{\pr a} = 2 \pi n,n \in \mathbb{N}
\eeq
Comparing this equation with the ZJJ quantization (\ref{eq:zj_sin}) for $\phi=2 \pi$, we obtain the following identification between $a^D$ and $A$:

\beq
A+2 \log \l \l \cfrac{2}{g} \r^{-B} \cfrac{2}{\Gamma(1/2-B)} \r - 2 \log \l \cfrac{2}{\sqrt{2 \pi} } \pm \sqrt{\cfrac{2}{\pi} - (-1)^B \cfrac{4}{\pi} \cos(\pi B) } \r = i a^D
\eeq
The choice between $+$ and $-$ in the second logarithm, as well as the value of $(-1)^B$ is the matter of analytic continuation from $g$ to $-g$
\footnote{After this text had appeared as a preprint, another paper \cite{basar} was published where authors made more precise identification between $A$ and $a^D$ using  small $g$ expansion. Actually, it turns out that
the relation (\ref{eq:simple}) holds in all orders in Plank constant - compare it with eq. (3.32) in \cite{basar}}. 
Fortunately, these terms vanish if we differentiate with respect to $g$ keeping $B$ constant. 
Performing the differentiation, we again arrive at eq. (\ref{eq:compar}).
Therefore, if we assume that $F^{Nek}=F^{WKB}$ we can actually prove the Dunne-\"Unsal relation.

Moreover, we claim that the Dunne-\"Unsal relation holds for every genus one potential. For higher genera exact quantization condition has not even been
conjectured yet. However the Whitham equations are the same so we can conjecture  that they play the role
of Dunne-Ünsal relations again. Note that we have used the Whitham dynamics for Riemann surfaces, that is for
holomorphic dynamical systems. However we could use the real version described above as well. In this case
the appropriate technique for the multi-regions in the phase space has been developed in \cite{multi}. 
We hope to consider the higher genus potentials  elsewhere.


\section{Conclusion}
In this paper we make some observations concerning  properties of the
complex Hamiltonian systems. We have argued that the AD point can be considered as the fixed
point from the Whitham dynamics viewpoint and it was shown that
anomalous dimensions at AD point  coincide with the Berry indexes in  the classical mechanics. Also, we have defined a "correlation length" for the mechanical system
near the AD point. We have derived Whitham equations for the $\Om$-deformed theory. Moreover we have
made the useful observation that the Dunne-\"Unsal relation  relevant for the exact quantization condition
can be considered as the equation of motion in the Whitham dynamics.

Certainly there is a lot to be done to treat the complex Hamiltonian systems
properly both classically and quantum mechanically.
In particular it would be important to clarify the
fate of the  Whitham hierarchy in the case of non-zero $\ep_1, \ep_2$
and develop its own quantization. It seems that this issue has
a lot in common with the generalization of the classical-quantum duality
from \cite{mvt, gzz} to the quantum-quantum case.


The work of A.G. and A.M. was supported in part
by grants RFBR-12-02-00284 and PICS-12-02-91052. The
work of A.M. was also supported by the Dynasty fellowship program.
A.G. thanks SCGP at Stony Brook University during the Simons Summer Workshop  where the part of this work has been done
for the hospitality and support.
We would like to thank G.~Basar, K.~Bulycheva, A.~Kamenev, G.~Korchemsky, P.~Koroteev, A.~Marshakov, A.~Mironov, A.~Morozov, A.~Neitzke and B.~Runov
for the useful discussions and comments. We are especially grateful to I.~Krichever for the careful reading of our manuscript.


\section{Appendix. Generalized Bethe ansatz from the Seiberg-Witten theory}
 \label{sec:bethe}
In this section we will consider Seiberg-Witten theory with the gauge group $SU(N_c)$ with $N_f$ fundamental matter hypermultiplets in the NS limit of $\Omega$ deformation. We will switch on
higher Whitham times and explicitly show how they deform spectral curve and Baxter equation.

Without higher Whitham times and $\Omega$-deformation, the case of $N_f=2 N_c$ corresponds
to the XXX spin chain with twist $h=\cfrac{-2q}{q+1}, q=\exp(2 \pi i \tau_{uv})$ and inhomogeneities $\th_l, J_l$. The spectral
curve reads as \cite{sw_int_rev}:
\begin{equation}
\label{eq:xxx_sc}
 - h A(x) w + (h+2)\cfrac{D(x)}{w} = 2 T(x)
\end{equation}
where $A(x), D(x), t(x)$ are the following polynomials:
\begin{equation}
 A(x)=\prod_{k=1}^{N_f}(x-\th_k- i J_k)
\end{equation}
\begin{equation}
 D(x)=\prod_{k=1}^{N_f}(x-\th_k+ i J_k)
\end{equation}
\begin{equation}
 T(x) = <\det(x-\phi)>=x^{N_c}-u_2 x^{N_c-2}+\dots
\end{equation}

Note that $q$ corresponds to ultraviolet coupling, $S$ and $T$ act as
\begin{eqnarray}
\label{eq:q_mod}
 S: q \ra 1-q \\ \nonumber
 T: q \ra \cfrac{q}{1-q}
\end{eqnarray}

Masses of hypermultiplets correspond to parameters
\begin{equation}
  m^F_k = \th_k- i J_k, \  m^{AF}_k = \th_k+ i J_k
\end{equation}
In the hyperelliptic parametrization the curve looks as
\begin{equation}
 y^2=T(x)^2+h(h+2)A(x)D(x)
\end{equation}
NS limit $\ep_1=0,\ \ep_2=\ep$ corresponds to the quantization of the XXX chain. Spectral curve (\ref{eq:xxx_sc}) promotes to the Baxter equation,
since $w$ becomes operator $w=\exp(i \ep \pr_x)$:
\begin{equation}
\label{eq:bax}
 -h A(x) Q(x+ i \ep) + (h+2) D(x) Q(x-i \ep) = 2 T(x) Q(x)
\end{equation}

The case of $N_f<2 N_c$ can be obtained by taking some of the masses to infinity, while keeping the product
\begin{equation}
 \La^{2 N_f} = m^F_1 \dots m^F_{N_f} m^{AF}_1 \dots m^{AF}_{N_f} q
\end{equation}
constant. It leads to the following spectral curve
\begin{equation}
 \La^{N_f} w + \cfrac{A_{N_f}(x)}{w} =  T(x)
\end{equation}
with $A_{N_f}(x)=\prod_{k=1}^{N_f}(x-m_k)$

Algebraic Bethe ansatz equations can be obtained by looking for the polynomial solution to the Baxter equation (\ref{eq:bax})
\begin{equation}
 Q(x)=(x-x_1)\dots (x-x_M)
\end{equation}
$M$ - is a magnon number, $x_k$ - Bethe roots.

Now, we consider non-zero Whitham times, which are coupling constants for the single-trace $N=2$ vector superfields (see eq. (\ref{eq:higher})).
Our considerations are close to those in \cite{bethe_xxx,deform}.

Nekrasov instanton partition function is equal to \cite{nek2002,nek2003}:
\begin{eqnarray}
 Z_{inst}= \sum_{\vec{Y}} q^{|\vec{Y}|} Z_{vec}(\vec{Y}) \prod_{n=1}^{N_f} Z_{hyp}(\vec{Y},m_n) \\ \nonumber
 Z_{vec}(\vec{Y}) = \prod_{(li)\neq (nj)} \cfrac{\Gamma(\ep_2^{-1}(x_{li}-x_{nj}-\ep_1))}{\Gamma(\ep_2^{-1}(x_{li}-x_{nj}))} \cfrac{\Gamma(\ep_2^{-1}(x_{li}^0-x_{nj}^0))}{\Gamma(\ep_2^{-1}(x_{li}^0-x_{nj}^0-\ep_1))} \\ \nonumber
 Z_{hyp}(\vec{Y},m) = \prod_{(li)} \cfrac{\Gamma(\ep_2^{-1}(x_{li}+m))}{\Gamma(\ep_2^{-1}(x_{li}^0+m))}
\end{eqnarray}
where $\vec{Y}={Y_1,...,Y_{N_c}}$ - set of Young diagrams, and $x_{li}, x_{li}^0$:
\begin{eqnarray}
 x_{li} = a_l +(i-1)\ep_1 + \ep_2 k_{li} \\ \nonumber
 x_{li}^0 = a_l + (i-1) \ep_1
\end{eqnarray}
$k_{li}$ - is the length of the i'th row in the diagram $Y_l$.

Let us denote by $t(x)$ the generating function of the Whitham times:
\begin{equation}
 t(x)= \sum_{k=1} T_k \cfrac{x^{k+1}}{k+1}
\end{equation}
Then the partition function is modified by the factor \cite{nek2003,marshakov07}:
\begin{equation}
\label{eq:u}
 U=\exp \l \cfrac{1}{\ep_1 \ep_2} \sum_{li} \l t(x_{li})+ t(x_{li}+\ep_1+\ep_2)-t(x_{li}+\ep_1)-t(x_{li}+\ep_2) \r \r
\end{equation}
In the NS limit, the sum over $\ep_2 k_{li} = y_{li}$ becomes continuous and we can consider it as an integral. Besides, we can use Stirling approximation
for the gamma functions $\Gamma(x) \approx \exp(x \log(x)-x) = \exp(f(x))$. Also, difference in (\ref{eq:u}) becomes derivative. After trivial manipulations:
\begin{eqnarray}
 Z_{inst}^{T} = \int \prod_{li} dy_{li} \exp(\cfrac{1}{\ep_2} H_{inst}^{T}(y)) \  \\ \nonumber
 H_{inst}^{T}(y) = V(x_{li}) - V(x_{li}^0) + \cfrac{\ep_2}{\ep_1}\sum_{li}( t'(x_{li}+\ep_1) - t'(x_{li})) \\ \nonumber
 V(x)= \log(q) \sum_{li} x_{li} + \sum_{li,n} f(x_{li}+m_n) + \cfrac{1}{2} \sum_{(li) \neq (nj)} (f(x_{li}-x_{nj}-\ep_1)-f(x_{li}-x_{nj}+\ep_1))
\end{eqnarray}
Integral over $y_{li}$ could be analyzed using saddle point method. Note, that all sums over $(li)$ become integrals over intervals $[x_{li}^0,x_{li}^0+y_{li}^{crit} ]$.
Let us introduce density function $\rho(x)$ which is constant on these intervals and vanishes elsewhere. Apart from the term with higher Whitham times, we obtain
the same expression as in \cite{bethe_xxx}:
\begin{equation}
 H_{inst}^T[\rho] = -\cfrac{1}{2} \int \ dx \ dy \ \rho(x) G(x-y) \rho(y) + \int \ dx \rho(x) \log(q R(x)) + \cfrac{1}{\ep_1} \int \ dx \rho(x) (t'(x+\ep_1)-t'(x))
\end{equation}
where:
\begin{eqnarray}
 G(x) = \cfrac{d}{dx} \log \l \cfrac{x-\ep_1}{x+\ep_1} \r \\ \nonumber
 R(x) = \cfrac{A(x) D(x)}{P(x) P(x+\ep_1)} \\ \nonumber
 P(x) = \prod_{l=1}^{N_c} (x-a_l)
\end{eqnarray}

Since $\rho$ is constant, variation over $y_{li}$ can be thought of as a variation of $\rho$.  Therefore, we end up with the following
saddle point equation:
\begin{equation}
 \cfrac{Q(x_{li}+\ep_1) Q^0(x_{li}-\ep_1)}{Q(x_{li}-\ep_1) Q^0(x_{li}+\ep_1)}=-qR(x_{li}) \exp(\cfrac{t'(x+\ep_1)-t'(x)}{\ep_1})
\end{equation}
where
\begin{equation}
\label{eq:qs}
 Q(x)=\prod_{l=1}^{N_c} \prod_{i=1}^\infty (x-x_{li}), \ \ Q^0(x)=\prod_{l=1}^{N_c} \prod_{i=1}^\infty (x-x_{li}^0)
\end{equation}
or using the explicit expression for the $x_{li}^0$:
\begin{equation}
 \cfrac{Q(x_{li}+\ep_1)}{Q(x_{li}-\ep_1)}=-qA(x_{li}) D(x_{li}) \exp \l \cfrac{t'(x+\ep_1)-t'(x)}{\ep_1} \r
\end{equation}
Indeed, we see that $T_1$ is responsible only for the shift of $\tau_{uv}$. This is the generalized Bethe ansatz equation we have mentioned before and one
can derive the following Baxter equation:
\begin{equation}
\label{eq:bax_high}
 - h \exp \l \cfrac{t'(x+\ep_1)}{\ep_1} \r A(x) Q(x+\ep_1) + (2+h) \exp \l \cfrac{-t'(x)}{\ep_1} \r D(x) Q(x-\ep_1) = 2 T(x) Q(x)
\end{equation}
In the classical limit $\ep_1 \ra 0$, the spectral curve reads as
\begin{equation}
 y^2=T(x)^2+(h+2)h A(x)D(x) \exp(t''(x))
\end{equation}
Several comments are in order. First of all, note that in (\ref{eq:qs}) products are infinite. It was argued  in \cite{bethe_xxx}, that if the following
quantization condition is imposed
\begin{equation}
\label{eq:xxx_quant}
 a_l = m_l - \ep_1 n_l, n_l \in \mathbb{Z},\ n_l > 0
\end{equation}
the most of the factors decouple
\begin{equation}
 x_{li}=x_{li}^0=a_l+(i-1) \ep , i \ge n_l
\end{equation}
and we are left with the polynomial Baxter function, that is with the algebraic Bethe ansatz.

However, it is apparent from the (\ref{eq:bax_high}) that $Q$ could not be polynomial because of the exponential factors. Nonetheless, we can get rid of them
by looking for a solution in the form
\begin{equation}
 Q(x)=F(x) \exp(C(x)/\ep_1)
\end{equation}
 where $F(x),C(x)$-polynomials. For $C(x)$ we have the following equations
\begin{eqnarray}
 t'(x+\ep_1)+C(x+\ep_1)-C(x)=0 \\ \nonumber
 -t'(x)+C(x-\ep_1)-C(x)=0
\end{eqnarray}
which are dependent. Therefore, we can always construct $C(x)$ from $t(x)$ unambiguously. For $F(x)$ we have the standard algebraic Bethe ansatz equations. One can repeat all
considerations from the \cite{bethe_xxx} and a that the quantization condition (\ref{eq:xxx_quant}) is not modified.

\bibliography{ref}

\end{document}